# Measurement of track structure parameters of low and medium energy helium and carbon ions in nanometric volumes

G. Hilgers[1], M.U. Bug[1], H. Rabus[1]

[1) *Physikalisch-Technische Bundesanstalt (PTB), Braunschweig, Germany*

**Abstract**

Ionisation cluster size distributions produced in the sensitive volume of an ion-counting wall-less nanodosimeter by monoenergetic carbon ions with energies between 45 MeV and 150 MeV were measured at the TANDEM-ALPI ion accelerator facility complex of the LNL-INFN in Legnaro. Those produced by monoenergetic helium ions with energies between 2 MeV and 20 MeV were measured at the accelerator facilities of PTB and with a $^{241}$Am alpha particle source. $C_3H_8$ was used as the target gas. The ionisation cluster size distributions were measured in narrow beam geometry with the primary beam passing the target volume at specified distances from its centre, and in broad beam geometry with a fan-like primary beam. By applying a suitable drift time window, the effective size of the target volume was adjusted to match the size of a DNA segment. The measured data were compared with the results of simulations obtained with the PTB Monte Carlo code PTra. Before the comparison, the simulated cluster size distributions were corrected with respect to the background of additional ionisations produced in the transport system of the ionised target gas molecules. Measured and simulated characteristics of the particle track structure are in good agreement for both types of primary particles and for both types of the irradiation geometry.

**Introduction**

Nanodosimetry focuses on investigating the physical characteristics of the microscopic structure of ionising particle tracks, i.e. the sequence of the interaction types and interaction sites of a primary particle and all its secondaries, which reflects the stochastic nature of the radiation interaction. Taking the particle track structure into account is particularly important for the biological effects of ion beams, where the major fraction of radiation damage is mainly concentrated along and in close vicinity to the primary particle trajectory.

In view of the emerging radiation therapy with carbon ions, manifesting itself in the emergence of carbon ion therapy facilities [1] [2] [3] and in the increasing number of treated patients [4], the ionisation structure of carbon ion tracks is of particular interest. This especially applies to the distal edge of the spread-out Bragg peak, which is expected to be of high importance for side effects of the treatment due to the pronounced increase of the RBE [5] [6]. Carbon ion therapy is, in fact, the therapy of choice for tumours close to critical tissues, because of the well-defined carbon ion range. However, critical tissues are at risk of being severely damaged due to the distal edge of the spread-out Bragg peak, which occupies the last 1−2 mm of the irradiated volume. The TANDEM-ALPI accelerator complex of the LNL-INFN can supply carbon ion beams of less than 270 MeV, having a range in tissue of ~1.7 mm. The energies of carbon ion beams used in this investigation are in the range between 40 MeV and 150 MeV, corresponding to a range in tissue from ~80 μm to ~650 μm [7]. Therefore, carbon ion beams of these energies are ideally suited to investigate the microscopic physical features of carbon ion beams at the distal edge of the spread-out Bragg peak.

Comparing the track structure of carbon ions with that of other light ions allows information on fundamental characteristics of the track structure of light ions to be obtained. Therefore, and also in view of the recent discussion [8] of using helium ions in radiation therapy [9], measurements with helium ions of 2 MeV and 20 MeV were also carried out at the accelerator facilities of PTB and with a $^{241}$Am alpha particle source. The range in tissue covered with helium ions of these energies lies between ~10 µm and ~350 µm [7].

**Nanodosimetric characteristics of particle track structure**

In nanodosimetry, the ionisation component of the particle track structure is of particular interest, and is characterised by the relative frequency distribution of the ionisation cluster size. The ionisation cluster size is defined as the number $v$ of ionisations generated in a target volume by a primary particle and its secondary electrons. As shown in Figure 1, often a cylindrical target volume is regarded for reasons of simplicity.

A primary particle of radiation quality $Q$ (where $Q$ is determined by the particle type and its energy) can either traverse the target volume or pass it at a distance $d$ (impact parameter) with respect to the longitudinal axis of the cylinder. The ionisation cluster size which is generated in the target can be interpreted as the superposition of the ionisation component of the particle track structure and of the geometric characteristics of the target volume. The ionisation cluster size distribution is characterised by the statistical distribution of the probabilities $P_v(Q,d)$ that exactly $v$ ions are created in the target volume. The probability distributions are normalised according to equation (1).

$$\sum_{v=0}^{\infty} P_v(Q,d) = 1 \qquad (1)$$

For the characterisation of the particle track, also the statistical moments of the probability distributions are suited which are calculated according to equation (2).

$$M_\xi(Q,d) = \sum_{\nu=0}^{\infty} \nu^\xi P_\nu(Q,d) \qquad (2)$$

with $\xi$ being the order of the moment of the distribution. Often, the first moment of the distribution, the mean ionisation cluster size $M_1(Q,d)$, which results from equation (3), is of particular interest.

$$M_1(Q,d) = \sum_{\nu=0}^{\infty} \nu \cdot P_\nu(Q,d) \qquad (3)$$

The ionisation cluster size distribution $P_\nu(Q,d)$ depends, on the one hand, on the radiation quality $Q$ and, on the other hand, on the geometry of the target volume and its material composition and density.

A subset of the ionisation cluster size distribution, which is of special interest, is the conditional cluster size distribution. It consists of those probabilities $P_\nu^C(Q,d)$ with ionisation cluster sizes of $\nu \geq 1$, i.e. it contains only those events, in which the primary particle has generated at least one ionisation in the target volume.

$$P_\nu^C(Q,d) = \frac{P_\nu(Q,d)}{\sum_{\nu=1}^{\infty} P_\nu(Q,d)} \quad \text{for} \ \nu \geq 1 \ \text{with} \ \sum_{\nu=1}^{\infty} P_\nu^C(Q,d) = 1 \qquad (4)$$

Consequently, the statistical moments of the conditional cluster size distribution are calculated according to equation (5).

$$M_\xi^C(Q,d) = \sum_{\nu=1}^{\infty} \nu^\xi P_\nu^C(Q,d) \qquad (5)$$

**Setup of the experiment**

The original setup of the experiment is described in detail in [10]. Later improvements regarding the data acquisition system and the data evaluation procedure as well as an improved characterisation of the device are described in detail in [11]. The nanodosimeter shown in Figure 2 consists of an interaction region filled with a rarefied target gas, an electrode system to extract ions from the interaction region, an evacuated acceleration stage with an ion-counting detector at its end and a primary particle detector.

The interaction region of the nanodosimeter, which is located between the electrodes of a plane parallel plate capacitor, is filled with the target gas at a pressure in the order of approximately 1 mbar. An ion entering the interaction region through the entrance aperture and traversing it parallel to the two electrodes is registered behind the exit aperture in a semiconductor detector. The ionised gas molecules generated by this primary particle and its secondaries drift towards the lower electrode due to the electrical field, which is applied across the plane parallel plate capacitor. Ions, which are generated in the interaction region's sensitive volume right above a small aperture in the lower electrode, are extracted from the ionisation region through the lower electrode. They are transported through ion optics to an ion-counting secondary electron multiplier, where they are detected. The shape and the size of the sensitive volume are defined by the spatial distribution of the detection efficiency of the generated ions, which is primarily determined by the electrical field strength as well as the working gas and its pressure. Repeating this measurement for a large number of single primary particles of radiation quality $Q$ at an impact parameter $d$, yields the relative frequency distribution $P_\nu(Q,d)$ of the ionisation cluster size $\nu$ of detected ions.

Figure 3 shows the general setup of the ion-counting nanodosimeter to carry out measurements at an ion accelerator. To reduce the primary particle rate, the ion beam hits an exchangeable thin gold foil in a scattering chamber attached in front of the nanodosimeter. Due to Rutherford scattering at the gold foil, a small number of ions is scattered horizontally towards the entrance aperture of the nanodosimeter or onto a semiconductor detector which serves as a monitor for the spectrometry of the scattered particles. For the measurements at the accelerator facilities of the LNL-INFN and PTB, the scattering chamber was attached to the beam line. The thickness of the gold foil depends on the energy of the primary carbon ions and ranged between 0.5 µm and 5.0 µm during the measurements at the LNL-INFN. In the measurements with helium ions at the PTB accelerators, a gold foil with a thickness of 0.1 µm was used.

A Mylar foil of 2.5 µm in thickness serves as an entrance window and is mounted on an aperture of 6 mm diameter, separating the high vacuum of the scattering chamber from the nanodosimeter. The energy losses of the primary ions in the gold foil and in the Mylar foil and, after passing the Mylar foil, in the target gas along the primary ion's track in the nanodosimeter were calculated with SRIM [7].

The data set with helium ions of 4 MeV was measured using alpha particles emitted from a $^{241}$Am source. Due to the coverage of the source's active area with 10 µm of Mylar, the energy spectrum of the emitted alpha particles, as measured with an alpha spectrometer, is shifted towards a mean energy of 4 MeV. In order to preserve the irradiation geometry, the source was positioned at the same distance to the target volume as the gold foil in the scattering chamber. The 2.5 µm Mylar foil was omitted.

In the setup described in [10] and [11], the nanodosimeter was only capable of measuring ionisation cluster size distributions for primary particles traversing the target volume centrally (impact parameter $d = 0$). In order to allow measurements of ionisation cluster size distributions for primary particles passing the target volume ($d \neq 0$), the trigger detector was replaced by a position-sensitive detector (PSD). The active area of the position-sensitive detector used for measurements with carbon ions is 20 mm in length and 3 mm in width (SiTek 1L20 [12]), and 10 mm in length and 2 mm in width for the measurements with helium ions (SiTek 1L10 [12]). The PSDs are not pixel based, but covered with a resistive layer and work according to the charge division principle. The impact position of the primary particle with respect to the centre of the detector is calculated according to equation (6):

$$X = L \cdot (Q_1 - Q_2)/(2 \cdot (Q_1 + Q_2)) \qquad (6)$$

Here $X$ is the distance from the detector centre, $L$ is the length of the detector, $Q_1$ and $Q_2$ are the charges collected at the two detector output terminals. The sum of both charges represents the energy of the primary particle.

The active area of the PSDs is not divided into well-defined separate pixels but is a "continuous" detection area, with the primary particle's position being determined by the ratio of the two charges collected at the output terminals. However, during data processing, virtual pixels of arbitrary size can be configured. Neglecting the position information results in the ionisation cluster size distribution of a broad beam radiation field, provided that the width of the nanodosimeter's target volume is substantially smaller than the width of the radiation field, as it is in the present setup.

Figure 4 shows the electrical configuration of the PSDs. The PSD itself is represented by the equivalent circuit diagram within the shaded box with $R_{PSD}$ symbolising the resistive layer of the PSD. The arrow leading from the anode of the diode D indicates a variable tap of $R_{PSD}$ and represents the position of the charged particle impinging on the detector surface. According to the position of the primary particle, the charge collected in the diode D is divided by the two different branches of $R_{PSD}$ into two output charges $Q_1$ and $Q_2$ according to the ratio of the branches of $R_{PSD}$. The bias voltage $U_B$ is applied at the cathode of the diode. At the two output terminals, the two charges $Q_1$ and $Q_2$ are tapped. The network consisting of $R_i$ and $C_i$ is the typical way to configure the connection of an integrating charge-sensitive preamplifier. The sum of the output signals of the two charge-sensitive preamplifiers is used to trigger the data acquisition system.

The output waveforms of the preamplifiers are digitised and processed with the method described in [13]. However, in contrast to an ordinary non-position-sensitive semiconductor detector, the output waveforms of the charge-sensitive preamplifiers connected to the PSD are affected by a side effect, which must be taken into account when processing the waveform data digitally. The two capacitors $C_1$ and $C_2$ are charged with $Q_1$ and $Q_2$. After being charged with $Q_1$ and $Q_2$, $C_1$ and $C_2$ start to exchange their charges across $R_{PSD}$ in order to equalise the charge on each capacitor, which leads to a deformation of the exponentially decaying output voltage of charge-sensitive preamplifiers with a resistive feedback. However, the algorithm for processing the waveforms of resistive feedback charge-sensitive preamplifiers described in [13] requires an undisturbed waveform. Therefore, the decay time constant of the feedback of the preamplifiers must be substantially smaller than the time constant for the charge equalisation of the capacitors $C_1$ and $C_2$ across $R_{PSD}$ for the charge exchange process not to disturb the waveform of the exponentially decaying output voltage of charge-sensitive preamplifiers. In this setup, A250 amplifiers [14] were chosen, since these amplifiers allow full access to the resistive feedback circuit and allow choosing the input FET to match the detector capacitance. The decay time constant of the A250 preamplifiers was chosen to be between 6 µs and 10 µs.

**Effective size of the target volume**

In order to obtain a target volume of the size of a DNA segment, a time window of ±2.5 µs centred in the maximum of the drift time distribution of the ionised target gas molecules was applied. Figuratively, the maximum of the drift time distribution is associated with the height of the primary ion beam above the lower electrode. The resulting spatial distribution of the extraction efficiency [11] for 1.2 mbar $C_3H_8$ is shown in Figure 5. Additionally, the figure shows, for the same target gas and pressure, the comparison of the spatial distribution of the extraction efficiency for the complete drift time distribution together with the corresponding simulation results. In the measured data, the mean ionisation cluster size $M_1(d,h)$, and in the simulated data, the extraction efficiency $\eta(d,h)$ [11] integrated along the primary particle track, are plotted as a function of the impact parameter $d$ and the height $h$ of the primary particle track above the nominal beam height, which is the height in which the particle track is parallel to the plane of the lower electrode of the interaction region. $M_1(d,h)$ and $\eta(d,h)$ are normalised relative to the respective maximum. The simulations for both drift time windows agree well with the measured data. The experimental determination of the spatial distribution of the extraction efficiency will be the subject of another paper.

The effective size of the target volume for the drift time window of ±2.5 µs was determined using two different methods. On the one hand, the extraction efficiency $\eta(d,h)$ was integrated along a primary particle track passing the target volume centrally at the nominal beam height, i.e. $h = 0$ mm and $d = 0$ mm, to determine the effective diameter, and along the line at $d = 0$ mm between $h = \pm 7$ mm, to determine the effective height. The result of this determination was an effective height $H$ of 1.83 mm or 0.4 µg/cm$^2$ and an effective diameter $D$ of 0.72 mm or 0.155 µg/cm$^2$ for 1.2 mbar $C_3H_8$. On the other hand, simulations were carried out using the PTra Monte Carlo code [15] [16] to simulate ionisation cluster size distributions created in a cylindrical target volume of a specified diameter and height. The probability of counting an ionisation was unity inside the target cylinder and zero outside. These simulations were carried out for the radiation qualities investigated and, after correcting the simulations with respect to the background of secondary ionisations [11], they were compared to the measured ionisation cluster size distributions for the impact parameter $d = 0$. This comparison is shown in Figure 6 for the mean ionisation cluster sizes $M_1(D,H)$ with different combinations of $D$ and $H$ determined from the simulated cluster size distributions and from the measured distributions. For helium ions and for carbon ions of energies above 80 MeV, a reasonable agreement between the simulated data set, which was obtained for an effective height $H = 1.83$ mm and an effective diameter $D = 0.72$ mm, and the measured data is achieved. For lower carbon ion energies, larger deviations are observed, which may be attributed to counting losses of the ionised target gas molecules due to the large number of target gas ions arriving within the drift time window of $\pm 2.5$ µs.

Applying the scaling procedure described in [17], which has been verified experimentally in [18], using the ratio $(\rho \lambda_{ion})_{C_3H_8}/(\rho \lambda_{ion})_{H_2O} = 0.66$ with $\lambda_{ion}$ being the ionisation mean free path calculated from the ionisation cross sections for $C_3H_8$ and $H_2O$ as used in PTra [16], leads to an approximate effective size of the target volume with $D \approx 0.23$ µg/cm$^2$ and $H \approx 0.61$ µg/cm$^2$, corresponding to 2.3 nm and 6.1 nm, respectively, expressed in terms of liquid water, which is in the order of the diameter and the height of two convolutions of the DNA strand.

**Uncertainties**

The uncertainties encountered in the experiment are described in detail in [11]. Due to the upgrade of the nanodosimeter with the position-sensitive detector, two additional contributions have to be taken into account, which affect the uncertainty of the impact parameter $d$. These two parameters are the position resolution of the PSD and the linearity of the determination of the position of the particle impinging on the PSD's surface. In order to determine these two parameters, a grid was placed in front of the detector close to the detector's surface, having a slit width and a strip width of 1 mm each. Then the detector was irradiated in the same setup as was used during the measurements of the ionisation cluster size distributions. These measurements were carried out for carbon ions with energies of 43 MeV, 75 MeV, 88 MeV and 150 MeV and for helium ions with energies of 2 MeV and 4 MeV. Figure 7 shows the intensity distribution of the carbon ions of 88 MeV impinging on the detector surface in dependence of the length coordinate (x-coordinate) of the detector. The width of the virtual pixels is 20 μm. The slits and the strips of the grid in front of the detector are clearly visible. The change in the intensity from the left edge of the detector to the right edge is due to the angular dependence of the Rutherford scattering of the primary beam at the gold foil in the scattering chamber (see Figure 7).

The position resolution is defined as the spacing between the two points of measurement of 10% intensity and of 90% intensity in an intensity profile across a sharp edge. Numerically, the position resolution can be determined by the convolution of a rectangular distribution representing the grid and a Gaussian distribution representing the slope of the intensity profile across the sharp edge with the FWHM of the Gaussian distribution being identical to the numerical value of the position resolution. Figure 8 shows the measured intensity profiles of two slits of the grid together with the corresponding fitted profiles obtained as described above, one in the centre of the detector around −0.4 mm and another one close to the edge around 7.6 mm. Except for the intensity, the two profiles differ by their position resolution (i.e. the FWHM of the Gaussian), which is about 50 µm in the centre and about 85 µm in a position close to the detector edge. This behaviour of the position resolution slightly decreasing with increasing distance is found in all measured intensity profiles.

In order to obtain the overall position resolution for the intensity profile over the whole length of the detector surface, the number of pixels in which the number of counts were between 10% and 90% of the mean value of the number of counts obtained in the pixels in the corresponding plateau was determined and averaged over the whole intensity profile. However, only data between the leftmost and the rightmost intensity minimum were taken into account in order to omit those data points which might be disturbed due to edge effects of the detector.

The relative root mean square (rms) detector non-linearity for the position is determined as [19]

$$\delta = \frac{\sqrt{\langle (X_m - X_t)^2 \rangle}}{L} \qquad (7)$$

with $X_m$ and $X_t$ corresponding to the measured and the true coordinates of the edges of the slits of the grid, respectively, and $L$ denoting the length of the active area of the PSD. Since the position of the grid with respect to the detector surface was not precisely reproducible, the exact position of the grid relative to the detector was reconstructed from the measured intensity profile. Due to the irradiation geometry, a magnification factor of about 1.02 had to be taken into account in the measured intensity profile. Figure 9 shows the measured and reconstructed positions of the edges of the slits of the grid. In the data derived from the measured intensity profile, the intensity is set to "1", if the number of the counts in the pixel is larger than the mean value averaged over all pixels in the region behind the respective slit. Otherwise it is set to "0", with the region behind the slit extending between the centres of the two neighbouring strips. From the resulting (digital) intensity profile, the positions of the edges of the slits of the grid are determined. To obtain the reconstruction data set, an intensity profile of a grid having a slit width and a strip width of 1 mm each and being magnified by a factor of 1.02 was shifted in such a way that the positions of the edges of the slits of the grid coincide best with the measured intensity profile. The detector non-linearity was then calculated from these two data sets according to equation (7).

For all data sets measured with the larger of the two detectors, i.e. with carbon ions, the position resolution was determined to be 81 μm. Together with the detector non-linearity of 96 μm, the total uncertainty in the determination of the position where the primary particle hits the detector amounts to ±126 μm. Due to a magnification factor of 1.3, which is determined by the geometrical setup of the experiment, the uncertainty of the impact parameter $d$ results in ±97 μm for the measurements with carbon ions. For measurements with the smaller detector, i.e. with helium ions, position resolutions of 158 μm and 90 μm were found for helium ions with energies of 2 MeV and 4 MeV, respectively. Taking into account the detector non-linearity of 47 μm leads to a total uncertainty in the determination of the position where the primary particle hits the detector of ±165 μm and ±102 μm, respectively. This results in a total uncertainty of the impact parameter $d$ of ±127 μm and ±78 μm for helium ions with energies of 2 MeV and 4 MeV, respectively. As the position resolution degrades with a decreasing signal-to-noise ratio, i.e. with decreasing energy deposited by the primary particle in the detector's active layer, it can be assumed that the position resolution for 20 MeV helium ions is at least as good as for 4 MeV helium ions. For carbon ions, no significant variation of the position resolution was observed in the energy range under investigation. This leads to the assumption that the position resolution is saturated for these energies deposited in the detector's active layer but is limited by other electronic or detector characteristics.

**Results**

Figure 10 shows ionisation cluster size distributions measured in 1.2 mbar $C_3H_8$ with carbon ions of 88 MeV energy for different impact parameters $d\rho$, given in mass per area with $\rho$ being the density, and the distribution for the whole range of $d\rho$ covered by the PSD. Depending on $d\rho$, the cluster size distributions are of different shapes. At $d\rho = 0$ µg/cm$^2$ the primary ion hits the target volume centrally, i.e. the fraction of the particle trajectory which is inside the target volume is maximal. As the cluster size is proportional to the ratio $(D\rho)/(\lambda\rho)$, with $D$ being the diameter of the sensitive volume and $\lambda$ being the mean free path length for ionisation, the corresponding cluster size distribution shows the highest occurrence of large ionisation clusters and the highest value of $\nu$ for the peak maximum. With increasing $d\rho$, the fraction of the particle trajectory which is inside the target volume decreases. Consequently, the occurrence of large ionisation clusters decreases and the peak in the frequency distribution shifts towards smaller values of $\nu$ ($d\rho = 0.17$ µg/cm$^2$ and $d\rho = 0.34$ µg/cm$^2$). When $d\rho$ has increased to such an amount that no more primary ions pass through the target volume, ionisations are exclusively produced by secondary electrons ($d\rho \geq 0.51$ µg/cm$^2$) created by ions passing by outside the target volume. Due to the decreasing solid angle covered by the target volume as seen from the secondary electrons at their point of emission, the probability of occurrence for large cluster sizes $\nu$ decreases further while the probability of occurrence for small cluster sizes increases with increasing $d\rho$. The cluster size distribution for the range of impact parameters $d\rho = \pm 1.71$ µg/cm² was obtained without discrimination of $d\rho$ and represents the cluster size distribution for broad beam irradiation geometry. It can be described as the superposition of the cluster size distributions for the respective impact parameters. The frequency of the ionisation clusters is maximal at $\nu = 0$ and decreases monotonically with increasing cluster size $\nu$ for the distribution with $d\rho = \pm 1.71$ µg/cm². However, it differs from the cluster size distributions for large impact parameters by a plateau-like region in the range of ionisation cluster sizes between $\nu = 4$ and $\nu = 10$, which is due to those primary ions having trajectories penetrating the target volume. The contribution of these ions is nevertheless only small, since the majority of the primary ions pass by outside the target volume. On the other hand, this small fraction of ions with trajectories penetrating the target volume contributes significantly to the biological effect due to the large amount of interaction processes produced inside the target volume.

Figure 11 shows the mean ionisation cluster size $M_1(Q,d\rho)$ measured in 1.2 mbar $C_3H_8$ for carbon ions of different energy as a function of the impact parameter $d\rho$. The shape of the measured $M_1(Q,d\rho)$ in dependence on $d\rho$ reflects the findings of the previous discussion: at $d\rho = 0$ µg/cm², where the primary ion hits the target volume centrally and the fraction of its trajectory being inside the target volume is maximal, hence, also $M_1(Q,d\rho)$ is maximal. With increasing $d\rho$ the fraction of the trajectory inside the target volume decreases, and $M_1(Q,d\rho)$ does as well. When $d\rho$ is in the range of the borders of the target volume, the decrease of $M_1(Q,d\rho)$ is most pronounced. A further increase of $d\rho$ leads to no more primary ions passing through the target volume, and the ionisations are produced by secondary electrons only. Due to the decreasing solid angle covered by the target volume as seen from the secondary electrons at their point of emission, $M_1(Q,d\rho)$ decreases further with increasing $d\rho$. However, due to the cross section for ionisation, which decreases with increasing primary particle energy in the energy range under investigation, the measured $M_1(Q,d\rho)$ profiles are shifted with increasing primary particle energy towards lower values of $M_1(Q,d\rho)$.

Figure 12 shows the ratio of the mean ionisation cluster sizes $M_1(Q,d\rho)$ divided by the corresponding mean number of primary ionisations ($D_{eff}/\lambda_{ion}$) produced by the primary ion along the effective diameter $D_{eff}$ of the target volume. $\lambda_{ion}$ represents the mean free path length for primary ionisation processes of the primary particle and is inversely proportional to the ionisation cross section of the respective primary particle in the target gas. By dividing by ($D_{eff}/\lambda_{ion}$), the differences between the $M_1(Q,d\rho)$ profiles corresponding to the different ion energies are reduced so that the measured $M_1(Q,d\rho)$ profiles almost coincide, especially for large impact parameters $d\rho$. Furthermore, it is found that for a central passage of the primary ion through the target volume, i.e. at $d\rho = 0$ µg/cm², the mean cluster size is close to $M_1(Q,d\rho) = 1$ for all $M_1(Q,d\rho)$ profiles. This behaviour confirms that the mean ionisation cluster size produced by primary ions in nanometric volumes is mainly determined by the proportionality to the ionisation mean free path length, whereas the contribution from secondary electrons, especially for large impact parameters $d\rho$, is almost invariant with the particle energy [20], at least in the range of energies investigated.

The ionisation cluster size distributions for the broad beam geometry with the impact parameter ranging between $d\rho = \pm 1.37$ µg/cm² measured for helium ions and between $d\rho = \pm 1.71$ µg/cm² for carbon ions of three different energies each are shown in Figure 13.

Independent of the energy and of the type of the primary ion, the frequency distributions are of similar shape. They mainly differ in the length of the plateau-like region, which is most pronounced for the ion of the respective type having the lowest energy (43 MeV for carbon ions and 2 MeV for helium ions) and only moderately visible for carbon ions of 150 MeV and even invisible for helium ions of 20 MeV. As discussed previously, the plateau is due to those primary ions having trajectories penetrating the target volume. As the cross section for ionisation increases with decreasing primary ion energy, and consequently the mean free path length for ionisation decreases, the number of ionisations produced by primary ions passing through the target volume increases, thus leading to an elongation of the plateau region towards larger ionisation cluster sizes with decreasing energy. On the other hand, the mean free path length for ionisation of 20 MeV helium ions is large, leading to a small number of ionisations inside the target volume and, therefore, the plateau in the cluster size distribution for 20 MeV helium ions vanishes completely. Since those ions with trajectories penetrating the target volume contribute significantly to the biological effect due to the large amount of interaction processes produced inside the target volume, the biological effect is expected to increase with decreasing energy due to the increase of interaction processes, which is reflected by the increasing length of the plateau region.

Figure 14 shows conditional ionisation cluster size distributions measured in 1.2 mbar $C_3H_8$ with carbon ions of 88 MeV energy for different impact parameters $d\rho$. At first glance, the conditional cluster size distributions shown on the left of Figure 14 do not seem to differ much from the ionisation cluster size distributions shown in Figure 10 except for the frequency of occurrence of specific clusters. However, on closer inspection it is found that the frequency distributions for $d\rho \geq 0.51$ µg/cm$^2$ coincide within the experimental uncertainties (Figure 14 right). This behaviour should be reflected in the statistical moments of the conditional cluster size distributions. The conditional mean ionisation cluster size $M_1^C(Q,d\rho)$ is shown in the upper plot of Figure 15. Compared to the monotonic decrease with increasing $d\rho$ of the mean ionisation cluster size $M_1(Q,d\rho)$ found in Figure 11, the conditional mean ionisation cluster size $M_1^C(Q,d\rho)$ shows asymptotic behaviour: for $d\rho > 0.75$ µg/cm², $M_1^C(Q,d\rho)$ shows almost the same constant value independent of $d\rho$. Furthermore, this constant value is not only found for carbon ions of 88 MeV energy, but for carbon ions of all energies investigated, and moreover, also for helium ions of all energies investigated. In order to extend the comparison of the statistical moments, $M_2^C(Q,d\rho)$ (middle plot of Figure 15) and $M_3^C(Q,d\rho)$ (lower plot of Figure 15) were calculated from the conditional cluster size distributions. Both $M_2^C(Q,d\rho)$ and $M_3^C(Q,d\rho)$ show the same behaviour as $M_1^C(Q,d\rho)$: for $d\rho > 0.75$ µg/cm², $M_2^C(Q,d\rho)$ and $M_3^C(Q,d\rho)$ show almost the same constant value independent of $d\rho$ for all radiation qualities investigated. However, the scatter of $M_\xi^C(Q,d\rho)$ increases with increasing $\xi$, since the influence of large clusters, which have a low frequency of occurrence and therefore a larger statistical uncertainty, is more pronounced (see equation (5)).

As mentioned previously, at large distances from the primary ion trajectory, when the primary ion passes by outside the target volume, the ionisation of the target gas inside the target volume is exclusively due to secondary electrons. By not taking into account the ionisation clusters of size $\nu = 0$, the influence of the decreasing flux, due to the decreasing solid angle covered by the target volume as seen from the secondary electrons at their point of emission, is reduced and only the influence of the secondary electron spectrum is preserved. The invariance of $M_\xi^C(Q,d\rho)$ with the distance $d\rho$ and with the radiation quality $Q$ shows that, at large distances, the secondary electron spectrum changes only slightly with the distance between the target volume and the primary ion trajectory and with the radiation quality $Q$ [20], at least for those radiation qualities and the range of $d\rho$ investigated.

**Comparison with Monte Carlo simulations**

The measured ionisation cluster size distributions were compared with Monte Carlo simulations using the Monte Carlo code PTra [15] [16]. The simulated cluster size distributions were corrected with respect to the background of additional ionisations using the procedure described in [11]. In brief, the measured ionisation cluster size distributions contain a background which arises from secondary ions produced due to the scattering of extracted target gas ions onto the cone-shaped electrode within the ion transport optics downstream of the extraction aperture. To compare the measured and simulated data, the background was included in the simulated ionisation cluster size distributions. The background in the experimental data was determined using a model, which is based on two quantities: the probability $\varepsilon$ of an ionised target gas molecule to hit the cone-shaped electrode after passing the extraction aperture and the expectation value $\lambda$ of a Poisson distribution, representing secondary ions created by a single ionised target gas molecule hitting the cone-shaped electrode. The degree of agreement between the measured and simulated data was described by the quantity $R$, defined in [11]. Due to the different length of the drift time window used in this investigation as compared to the length of the drift time window used in [11], the values of $\varepsilon = 0.014$ and $\lambda = 9$ (see Figure 16) obtained for the common minimum of the minimum deviation between the measured and the simulated data for the combination of all data sets differ from the values for $\varepsilon = 0.0065$ and $\lambda = 15$ obtained in [11].

Figure 17 shows the comparison of the mean ionisation cluster sizes $M_1(Q,d\rho)$ obtained from measured and simulated background-corrected ionisation cluster size distributions for helium ions (left) and carbon ions (right), respectively, in 1.2 mbar $C_3H_8$. For better clarity, the mean cluster sizes for 75 MeV and 88 MeV carbon ions are omitted.

For carbon ions, the data show a generally good agreement between measurements and simulations. For impact parameters $d\rho = 0$ µg/cm², i.e. for a central passage of the primary ion through the target volume, and for $d\rho \geq 0.7$ µg/cm², only minor deviations between the measured and the simulated mean cluster sizes are found, with the simulated $M_1(Q,d\rho)$ being slightly larger than those measured, except for carbon ions of 43 MeV and $d\rho = 0$ µg/cm². Here, the measured $M_1(Q,d\rho)$ is apparently smaller than the simulated one, which might be attributed to counting losses in the secondary electron multiplier due to the large number of ionised target gas molecules. In the range of $d\rho$ between 0.15 µg/cm² $\leq d\rho \leq$ 0.55 µg/cm², where the fraction of the trajectory inside the target volume decreases down to a grazing passage at the edge of the target volume, the measured $M_1(Q,d\rho)$ are apparently larger than those simulated (again except for the 43 MeV carbon ions, which suffer from counting losses). These differences at the border of the target volume might be caused by imperfections in the shape of the simulated spatial distribution of the extraction efficiency, which enters into the Monte Carlo simulation of the ionisation clusters.

For helium ions of 2 MeV in the range of $d\rho$ between 0.25 µg/cm² $\leq d\rho \leq$ 0.75 µg/cm², the measured $M_1(Q,d\rho)$ differ significantly from those simulated. The reason for this difference is not clear. However, for helium ions of 4 MeV and 20 MeV, the degree of the agreement between the measurement and the simulation is similar to that found for carbon ions.

The comparison of measured and simulated background-corrected ionisation cluster size distributions is shown in Figure 18 for selected values of the impact parameters $d\rho$ for 94 MeV carbon ions in 1.2 mbar $C_3H_8$, and in Figure 19 for 4 MeV helium ions.

For $d\rho = 0$ µg/cm$^2$, significant differences between the measurement and the simulation for carbon ions are found for large cluster sizes $\nu \gtrsim 30$, whereas for cluster sizes around the maximum of the distribution, i.e. $\nu \lesssim 20$, the agreement is excellent. For the cluster size distributions of the other values of $d\rho$, only minor deviations are found in the comparison between the measurements and the background-corrected simulations.

For helium ions, the comparison between the measurement and the simulation shows significant deviations for $d\rho = 0.34$ µg/cm$^2$ for large cluster sizes $\nu \gtrsim 18$ and for cluster sizes in the range $5 \lesssim \nu \lesssim 10$. For the cluster size distributions of the other values of $d\rho$, the degree of the agreement between the measurement and the simulation is similar to that found for carbon ions.

Figure 20 shows the comparison of measured and simulated background-corrected ionisation cluster size distributions for broad beam geometry with the impact parameter ranging between $d\rho = \pm 1.37$ µg/cm$^2$ measured for helium ions (left) and between $d\rho = \pm 1.71$ µg/cm$^2$ for carbon ions (right) in 1.2 mbar $C_3H_8$.

For 150 MeV carbon ions, the frequency in the occurrence of clusters along the falling slope towards increasing cluster size is systematically smaller in the simulation as compared to the measured data. For 75 MeV carbon ions, an overall good agreement is found. Only in the knee at cluster size $\nu \cong 10$ are minor deviations found between the measurement and the simulation.

For 2 MeV helium ions, the frequency in the occurrence of cluster sizes in the range $15 \lesssim \nu \lesssim 20$ is lower in the simulation than in the measurement and larger for cluster sizes $\nu > 35$. For 20 MeV, the behaviour of the frequency in the occurrence of clusters is the opposite: here the occurrence of cluster sizes in the range $10 \lesssim \nu \lesssim 15$ is larger in the simulation than in the measurement and lower for cluster sizes $\nu > 20$.

In total, a good agreement is found in the comparison between measured and simulated background-corrected ionisation cluster size distributions for both types of irradiation conditions, i.e. for the narrow beam geometry with specified impact parameters $d\rho$ as well as for the broad beam geometry with the impact parameter ranging between $d\rho = \pm 1.37$ µg/cm² for helium ions and between $d\rho = \pm 1.71$ µg/cm² for carbon ions.

**Conclusions**

Ionisation cluster size distributions for helium and carbon ions were measured in a target volume having a cylindrical effective volume with a diameter of $D \approx 0.23$ µg/cm² and a height of $H \approx 0.61$ µg/cm², expressed in terms of liquid water. The effective size of the target volume corresponds to the size of a DNA segment. The ranges of the primary ions in tissue were from ~10 µm to ~650 µm and are located at the distal edge of the spread-out Bragg peak.

Depending on the impact parameter $d\rho$, the cluster size distributions are of different shapes, which are determined by the combined effect of the length of the fraction of the particle trajectory inside the target volume on the one hand, and by the solid angle covered by the target volume as seen from the secondary electrons at their point of emission, on the other hand. For small values of $d\rho$, where the fraction of the particle trajectory inside the target volume is large, the distributions show a peak at cluster sizes $\nu > 0$, whereas for larger values of $d\rho$, where the fraction of the particle trajectory inside the target volume is small, the frequency of the ionisation clusters is maximal at $\nu = 0$ and decreases monotonically with increasing cluster size $\nu$. When $d\rho$ has increased to such an amount that no more primary ions pass through the target volume, ionisations are exclusively produced by secondary electrons created by ions passing by outside the target volume. Due to the decreasing solid angle covered by the target volume as seen from the secondary electrons at their point of emission, the cluster size $\nu$ decreases further with increasing $d\rho$.

The cluster size distribution for broad beam irradiation geometry can be described as the superposition of the cluster size distributions for the respective impact parameters. The frequency of the ionisation clusters is maximal at $\nu = 0$ and decreases monotonically with increasing cluster size $\nu$. However, it shows a plateau-like region, which is due to primary ions penetrating the target volume. However, the contribution of these ions is only small, since the majority of the primary ions pass by outside the target volume. On the other hand, this small fraction of ions with trajectories penetrating the target volume contributes significantly to the biological effect due to the large amount of interactions inside the target volume.

The statistical moments of the conditional cluster size distributions $M_1^C(Q,d)$, $M_2^C(Q,d)$ and $M_3^C(Q,d)$ show asymptotic behaviour: for $d\rho > 0.75$ µg/cm², $M_\xi^C(Q,d)$ show almost the same constant value independent of $d\rho$. Furthermore, this same constant value is found for carbon ions of all energies investigated, and moreover, also for helium ions of all energies investigated. In large distances from the primary ion trajectory, when the primary ion passes by outside the target volume, the ionisation of the target gas inside the target volume is exclusively due to secondary electrons. By not taking into account the ionisation clusters of size $\nu = 0$, the influence of the decreasing flux, due to the decreasing solid angle covered by the target volume as seen from the secondary electrons at their point of emission, is reduced and only the influence of the secondary electron spectrum is preserved. The invariance of $M_\xi^C(Q,d)$ from the distance $d\rho$ and from the radiation quality $Q$ shows that, at large distances, the secondary electron spectrum changes only slightly with the distance between the target volume and the primary ion trajectory along with the radiation quality $Q$ [20], at least for those radiation qualities and the range of $d\rho$ investigated.

The measured ionisation cluster size distributions were compared with Monte Carlo simulations using the Monte Carlo code PTra [15] [16]. The simulated cluster size distributions were corrected with respect to the background of additional ionisations using the procedure described in [11]. In total, a good agreement is found in the comparison between measured and simulated background-corrected ionisation cluster size distributions for both types of irradiation conditions, i.e. for the narrow beam geometry with specified impact parameters $d\rho$ as well as for the broad beam geometry with the impact parameter ranging between $d\rho = \pm 1.37$ µg/cm² for helium ions and between $d\rho = \pm 1.71$ µg/cm² for carbon ions.

**Acknowledgement**

The authors gratefully acknowledge the developers of the device from the Weizmann Institute of Science, Rehovot, Israel, for transferring the nanodosimeter as described in [10] to PTB for further use. The authors would like to thank the staff of the ion accelerator facilities at the LNL-INFN and at PTB for their assistance and support during the measurements, and Dr. D. Moro (formerly at LNL-INFN) for his assistance and support in the preparation of the measurements. The authors also express their gratitude to W. Helms†, B. Lambertsen and A. Pausewang for their invaluable contributions to the preparation and carrying out of measurements and their assistance in data processing.

This work was partly funded within EMRP Joint Research Project SIB06 BioQuaRT and EC grant agreement 262010 (ENSAR). The EMRP is jointly funded by the EMRP participating countries within EURAMET and the European Union.

# References

[1] *S.E. Combs, O. Jäkel, T. Haberer, J. Debus*: Particle therapy at the Heidelberg Ion Therapy Center (HIT) – Integrated research-driven university-hospital-based radiation oncology service in Heidelberg, Germany. Radiother. Oncol. 95 – 1 (2010), 41 – 44.

[2] *M. Benedikt, A. Wrulich*: MedAustron – Project overview and status. Eur. Phys. J. Plus 126 (2011), DOI: 10.1140/epjp/i2011-11069-9.

[3] *S. Rossi*: The status of CNAO. Eur. Phys. J. Plus 126 (2011), DOI: 10.1140/epjp/i2011-11078-8.

[4] *M. Jermann*: Particle Therapy Statistics in 2014. Int. J. Particle Ther. 2 (2015); 50 – 54.

[5] *R. Grün, T. Friedrich, T. Elsässer, M. Krämer, K. Zink, C.P. Karger, M. Durante, R. Engenhart-Cabillic, M. Scholz*: Impact of enhancements in the local effect model (LEM) on the predicted RBE-weighted target dose distribution in carbon ion therapy. Phys. Med. Biol. 57 (2012), 7261 – 7274.

[6] *R. Grün, T. Friedrich, M. Krämer, K. Zink, M. Durante, R. Engenhart-Cabillic, M. Scholz*: Physical and biological factors determining the effective proton range. Med. Phys. 40 – 11 (2013), 111716-1.

[7] *J.F. Ziegler, M.D. Ziegler, J.P. Biersack*: SRIM – The Stopping and Range of Ions in Matter Version 2006.02, (2006), *URL*: http://www.srim.org/

[8] *M. Krämer, E. Scifoni, C. Schuy, M. Rovituso, W. Tinganelli, A. Maier, R. Kaderka, W. Kraft-Weyrather, S. Brons, T. Tessonnier, K. Parodi, M. Durante*: Helium ions for radiotherapy? Physical and biological verifications of a novel treatment modality. Med. Phys. 43(4) (2016) 1995 – 2004.

[9] *W. Saunders, J.R. Castro, G.T.Y. Chen, J.M. Collier, S.R. Zink, S. Pitluck, T.L. Phillips, D. Char, P. Gutin, G. Gauger, C.A. Tobias, E.L. Alpen*: Helium-Ion Radiation Therapy at the Lawrence Berkeley Laboratory: Recent Results of a Northern California Oncology Group Clinical Trial. Radiat. Res. 104 (1985) S-227 – S-234.

[10] *G. Garty, S. Shchemelinin, A. Breskin, R. Chechik, G. Assaf, I. Orion, V. Bashkirov, R. Schulte, B. Grosswendt*: The performance of a novel ion-counting nanodosimeter. NIM A 492 (2002), 212 - 235.

[11] *G. Hilgers, M. Bug, E. Gargioni, H. Rabus*: Secondary ionisations in a wall-less ion-counting nanodosimeter: quantitative analysis and the effect on the comparison of measured and simulated track structure parameters in nanometric volumes. Eur. Phys. J. D 69 (2015).

[12] *URL*: http://www.sitek.se/

[13] *V.T. Jordanov, G.F. Knoll, A.C. Huber, J.A. Pantazis*: Digital techniques for real time pulse shaping in radiation measurements, NIM A 353 (1994), 261 – 264.

[14] *URL*: http://amptek.com/

[15] *B. Grosswendt*: Formation of ionization clusters in nanometric structures of propane-based tissue-equivalent gas or liquid water by electrons and α-particles. Radiat. Environ. Biophys. 41 (2002), 103 – 112.

[16] *M.U. Bug, E. Gargioni, H. Nettelbeck, W. Y. Baek, G. Hilgers, A. B. Rosenfeld, H. Rabus*: Review of ionization cross section data of nitrogen, methane and propane for light ions and electrons and their suitability for use in track structure simulations, Phys. Rev. E 88 (2013), 043308.

[17] *B. Grosswendt*: Nanodosimetry, the metrological tool for connecting radiation physics with radiation biology, Radiat. Prot. Dosim. 122 (2006), 404 – 414.

[18] *G. Hilgers*: Check of the scaling procedure of track structures of ionizing radiation in nanometric volumes, Rad. Meas. 45 (2010), 1228 – 1232.

[19] *A. Banu, Y. Li, M. McCleskey, M. Bullough, S. Walsh, C.A. Gagliardi, L. Trache, R.E. Tribble, C. Wilburn*: Performance evaluation of position-sensitive silicon detectors with four-corner readout, NIM A 593 (2008), 399 – 406.

[20] V. Conte, P. Colautti, B. Grosswendt, D. Moro, L. De Nardo: Track structure of light ions: experiments and simulations. New J. of Phys. 14 (2012) 093010.

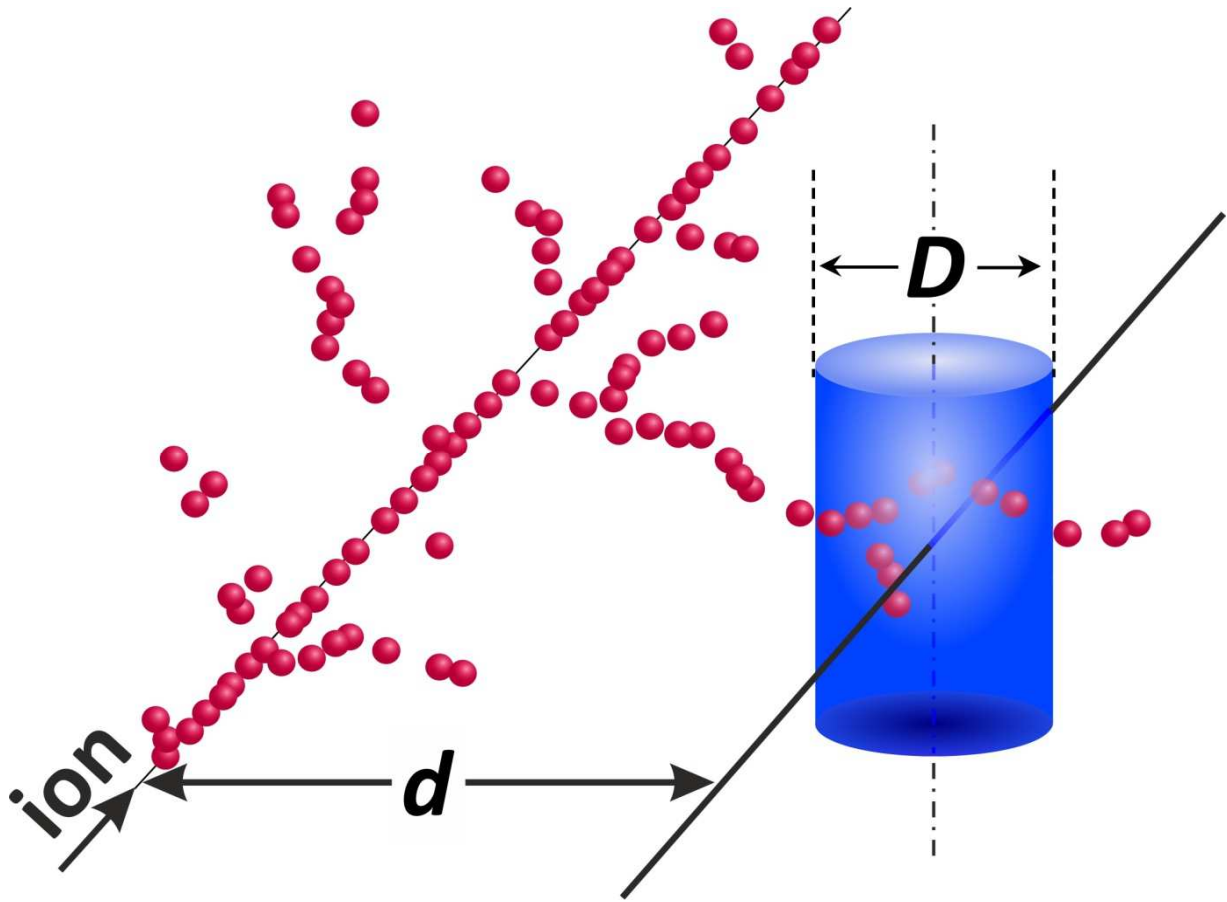

Figure 1: Schematic representation of the formation of an ionisation cluster by an ionising particle which passes by a cylindrical target volume of diameter $D$ at a distance $d$ from the cylinder axis. In the shown segment of the particle track, the solid circles represent the locations of ionisation interaction.

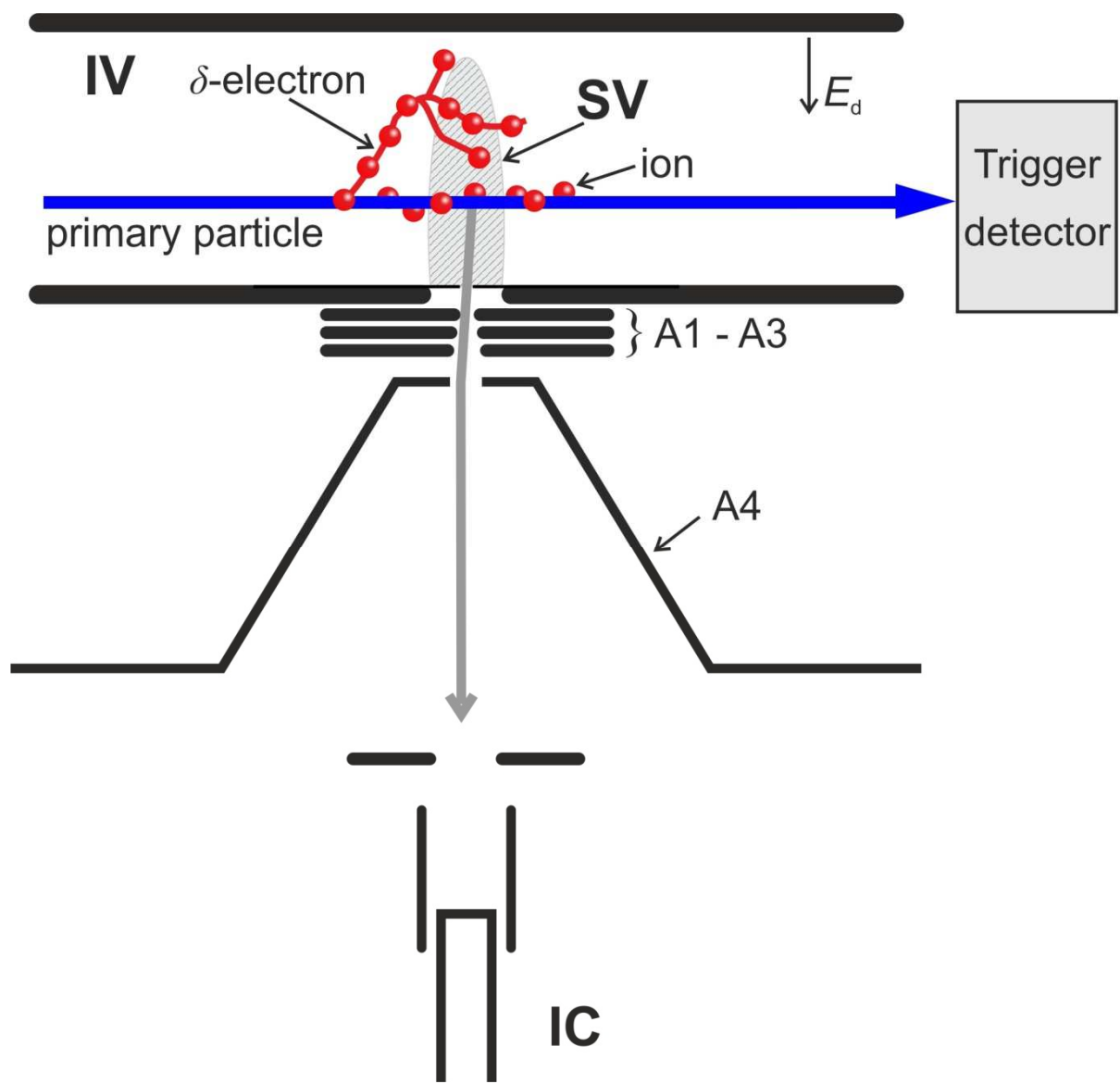

Figure 2: Schematic representation of the ion-counting nanodosimeter (adapted from [10]). A charged particle passing through the interaction volume (IV) creates ions in the target gas, which drift towards the bottom electrode due to the electric field $E_d$. Only ions created within the sensitive volume (SV) are extracted via a small aperture in the bottom electrode. These ions are then focused and accelerated via the electrodes A1 - A4 into the ion counter (IC), where charge pulses are generated by secondary electron multiplication. Note that both the SV and the δ-electron track are schematic representations and not to scale.

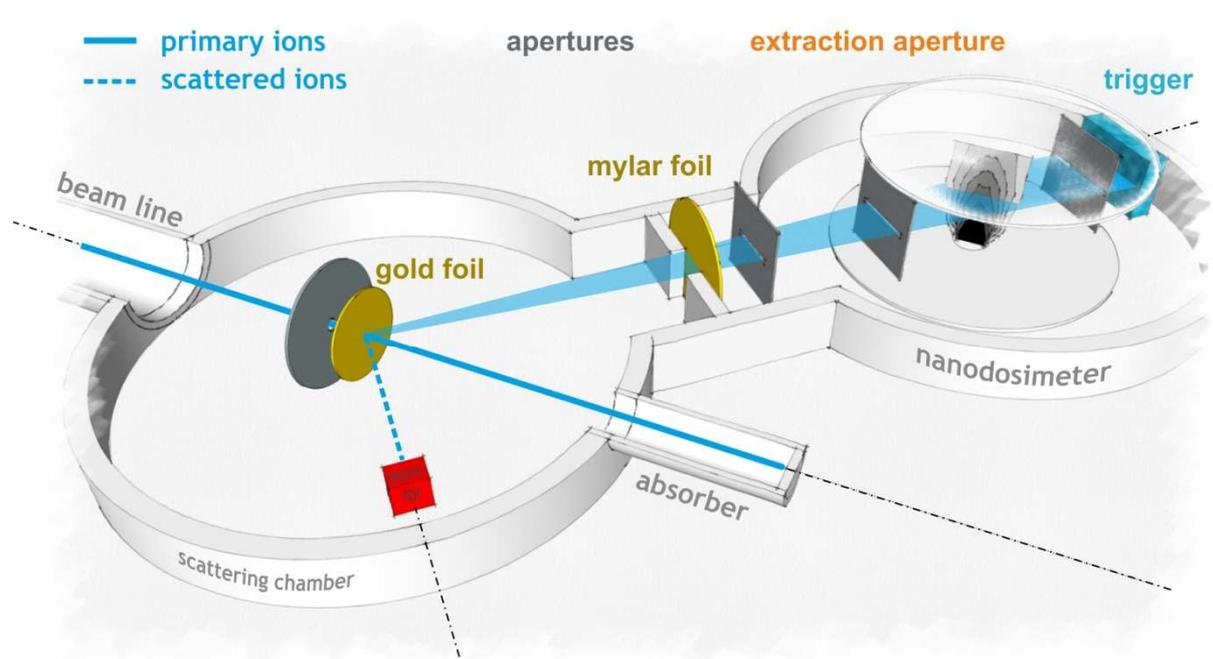

Figure 3: Schematic setup of the PTB ion-counting nanodosimeter for measurements at an ion accelerator.

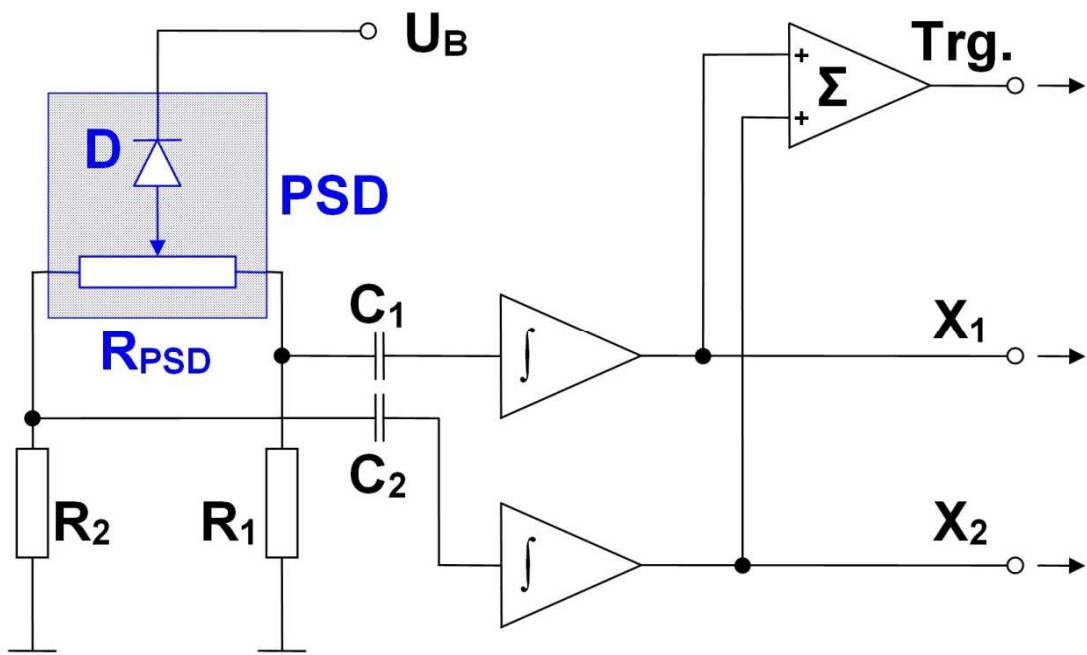

Figure 4: Electrical configuration of the PSDs. The PSD itself is represented by the equivalent circuit diagram within the shaded box with $R_{PSD}$ symbolising the resistive layer of the PSD. The arrow leading from the anode of the diode D indicates a variable tap of $R_{PSD}$ and represents the position of the charged particle impinging on the detector surface.

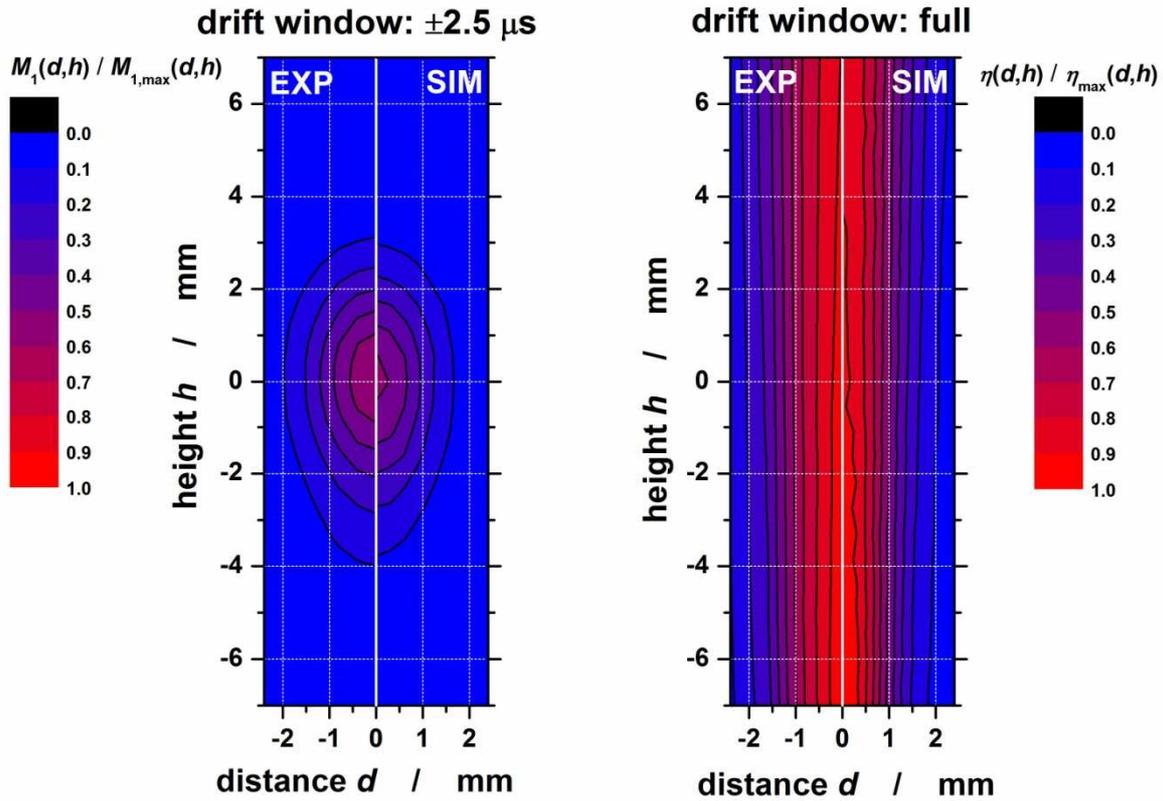

Figure 5: Spatial distribution of the extraction efficiency for 1.2 mbar $C_3H_8$ for the drift time window of ±2.5 μs centred in the maximum of the drift time distribution of the ionised target gas molecules (left) and the complete drift time distribution (right). Each plot shows the measured distribution for $d < 0$ and the simulated distribution for $d > 0$. For the measurements, the colours represent the mean ionisation cluster size normalised with respect to its maximum value. For the simulations, the colours represent the extraction efficiency normalised with respect to its maximum value.

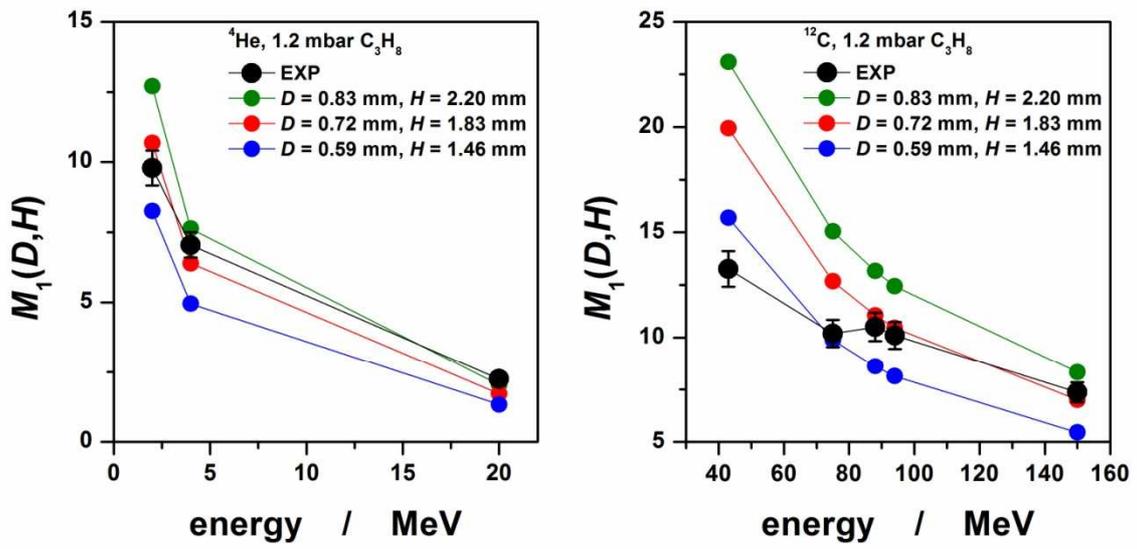

Figure 6: Mean ionisation cluster sizes $M_1(D,H)$ for different combinations of $D$ and $H$ determined from the simulated and the measured ionisation cluster size distributions for helium ions (left) and carbon ions (right).

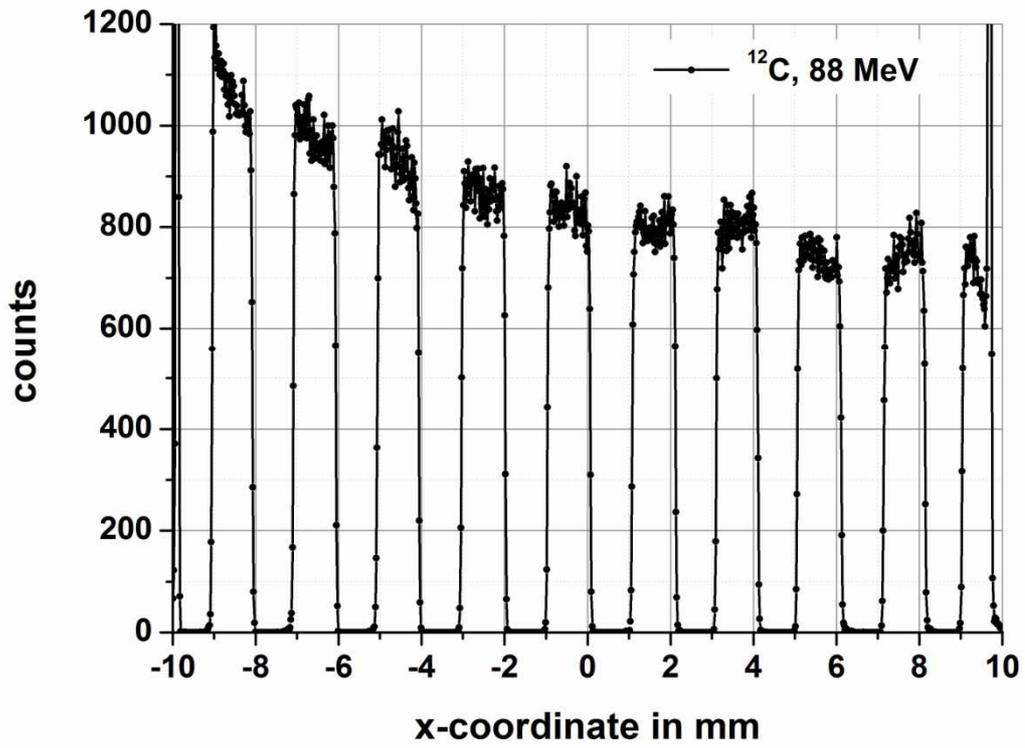

Figure 7: Intensity distribution of the carbon ions of 88 MeV impinging on the detector surface in dependence of the length coordinate (x-coordinate) of the detector. The width of the virtual pixels is 20 μm.

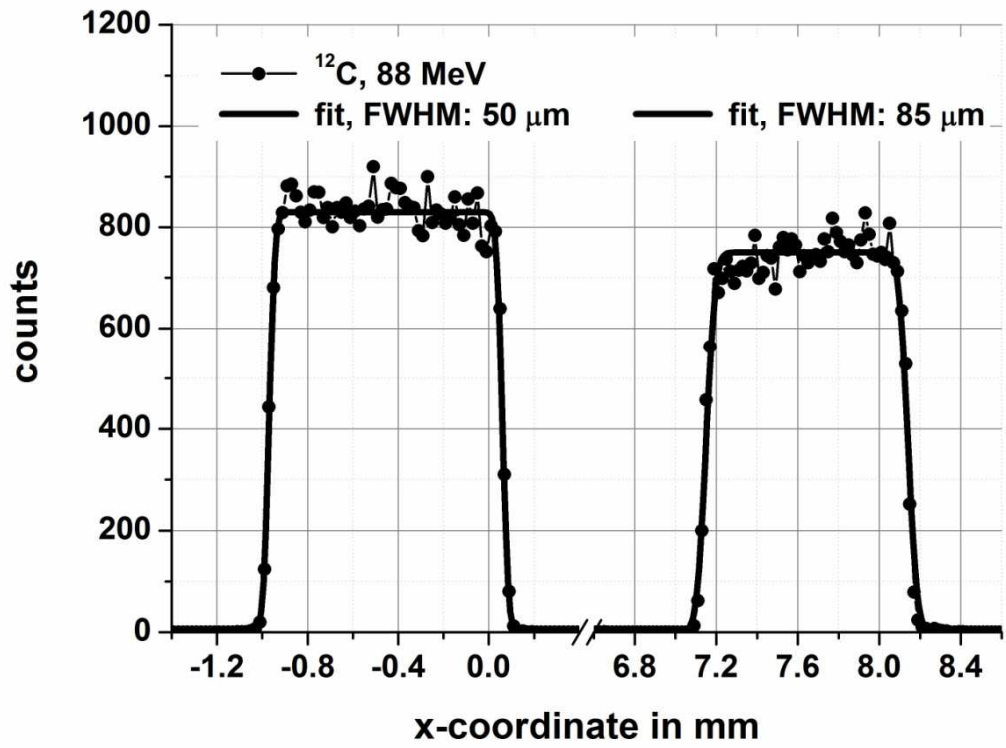

Figure 8: Measured intensity profiles of two slits of the grid together with the corresponding fitted profiles.

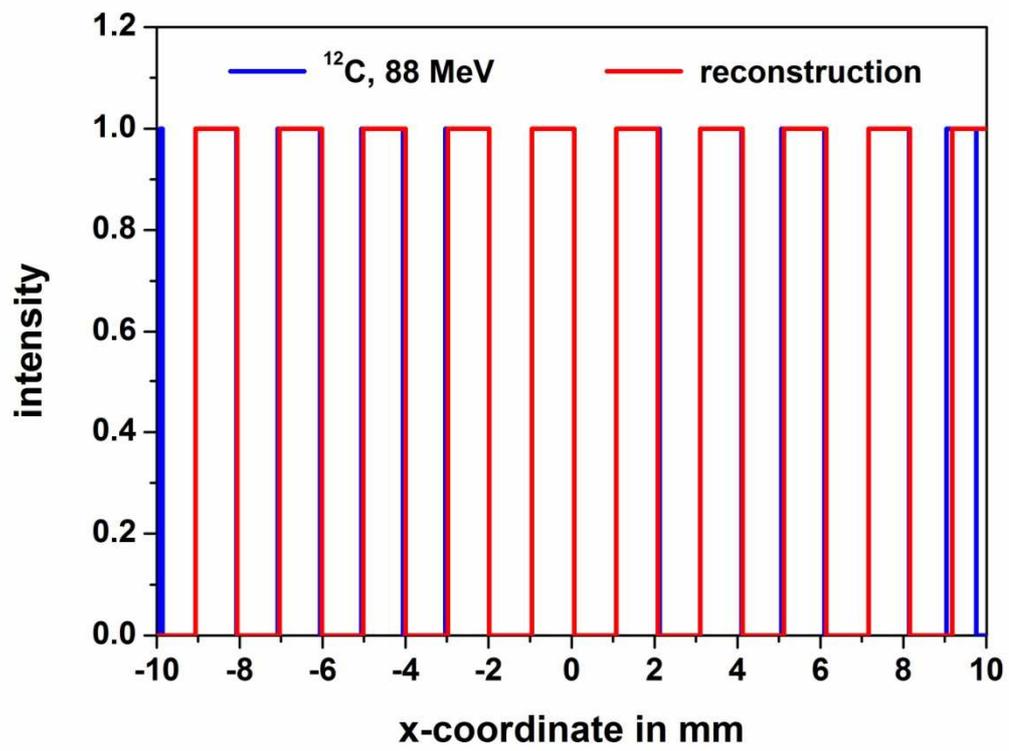

Figure 9: Measured and reconstructed positions of the edges of the slits of the grid.

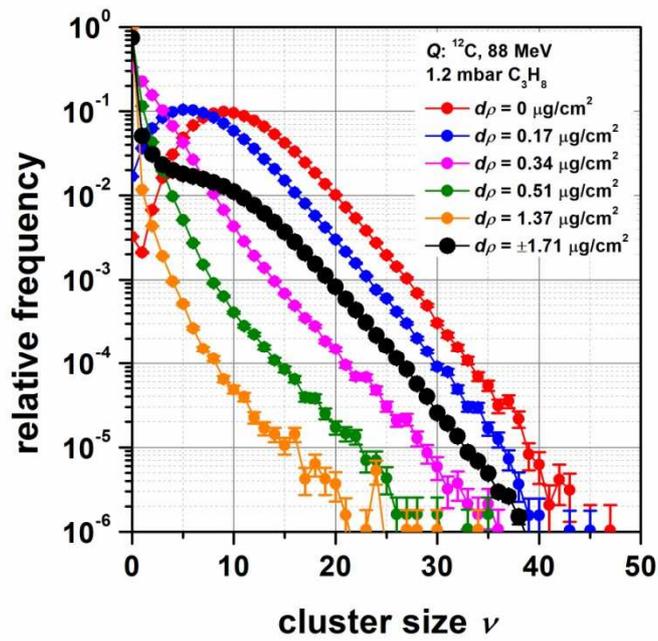

Figure 10: Ionisation cluster size distributions measured in 1.2 mbar $C_3H_8$ with carbon ions of 88 MeV energy for different impact parameters $d\rho$ and the distribution for the whole range of $d\rho$ covered by the PSD.

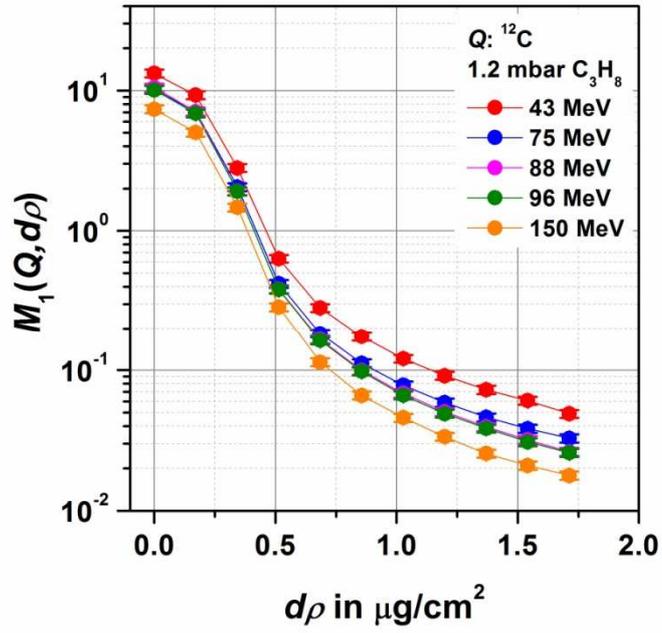

Figure 11: Mean ionisation cluster size $M_1(Q,d\rho)$ measured in 1.2 mbar $C_3H_8$ for carbon ions of different energy as a function of the impact parameter $d\rho$.

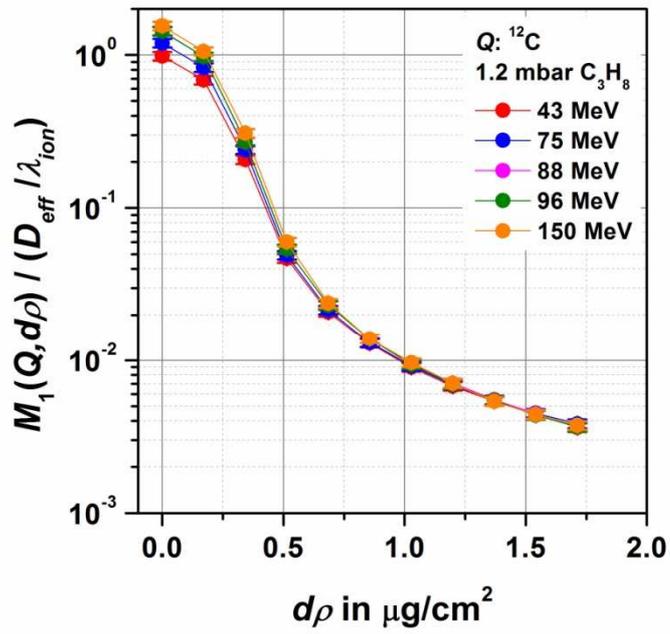

Figure 12: Ratio of the mean ionisation cluster sizes $M_1(Q,d\rho)$ divided by the corresponding mean number of primary ionisations ($D_{eff}/\lambda_{ion}$) produced by the primary ion.

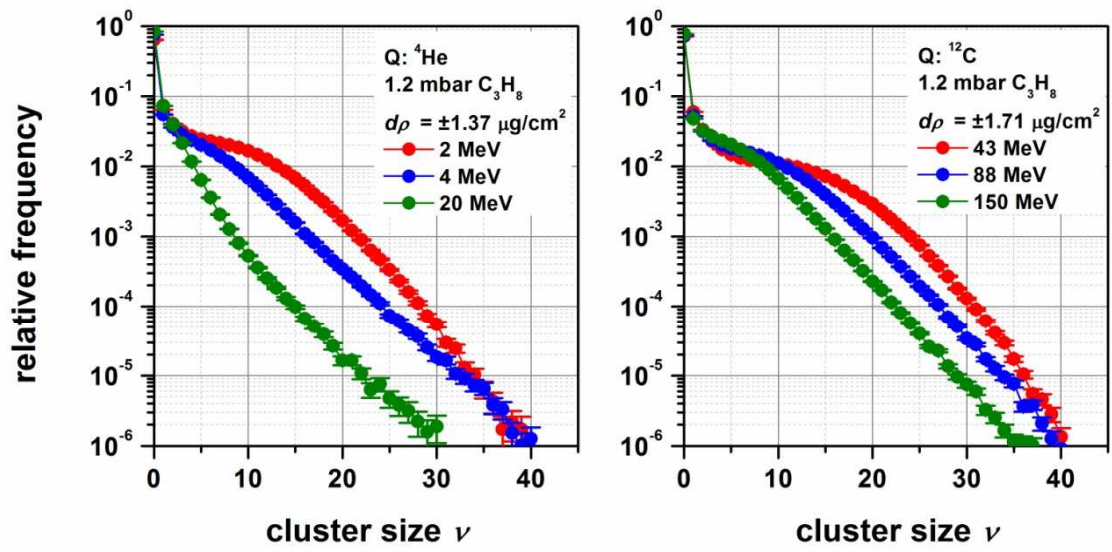

Figure 13: Ionisation cluster size distributions for the broad beam geometry with the impact parameter ranging between $d\rho = \pm 1.37$ μg/cm$^2$ measured for helium ions (left) and between $d\rho = \pm 1.71$ μg/cm$^2$ for carbon ions (right).

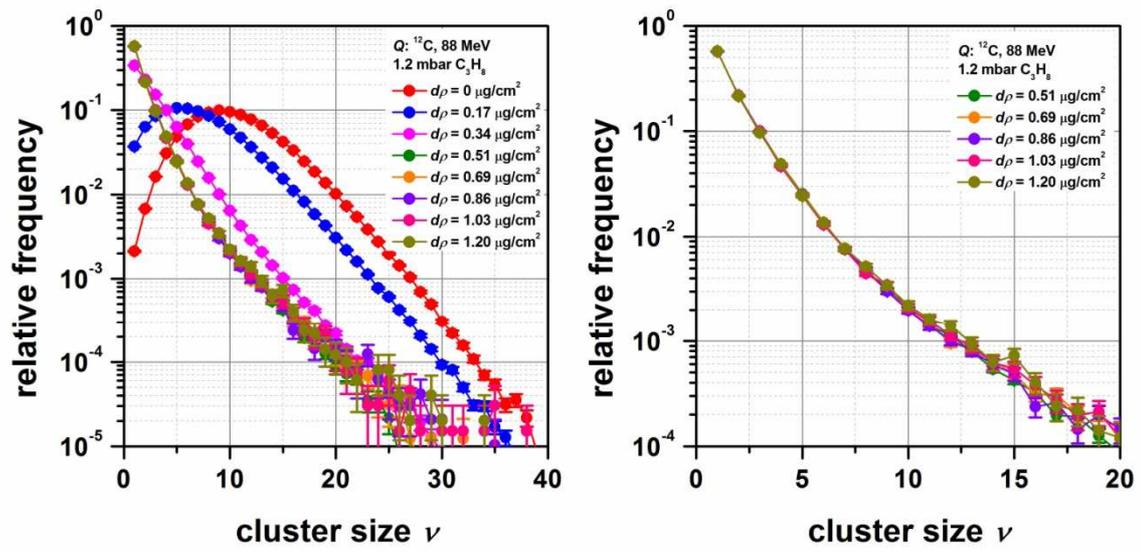

Figure 14: Conditional ionisation cluster size distributions measured in 1.2 mbar $C_3H_8$ with carbon ions of 88 MeV energy for different impact parameters $d\rho$. In the diagram on the right, an enlarged section of the data shown in the left diagram is plotted.

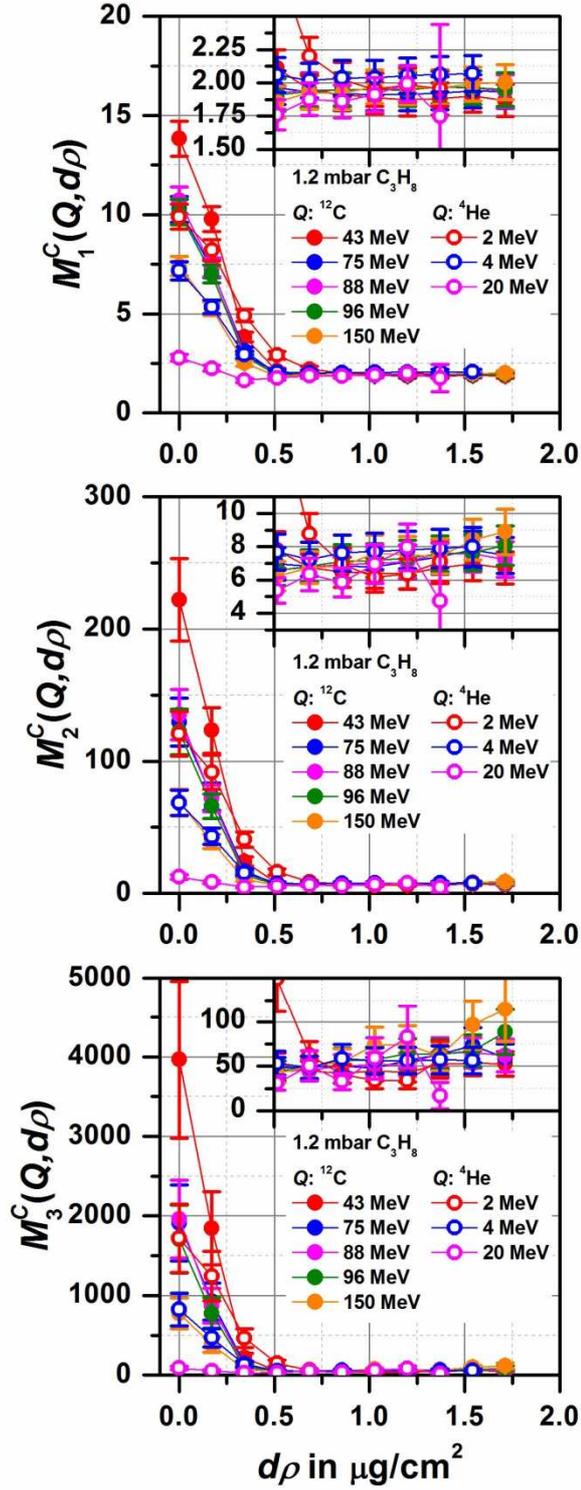

Figure 15: Statistical moments $M_\xi^C(Q,d\rho)$ of the conditional cluster size distributions measured in 1.2 mbar $C_3H_8$ with helium ions and carbon ions for different impact parameters $d\rho$.

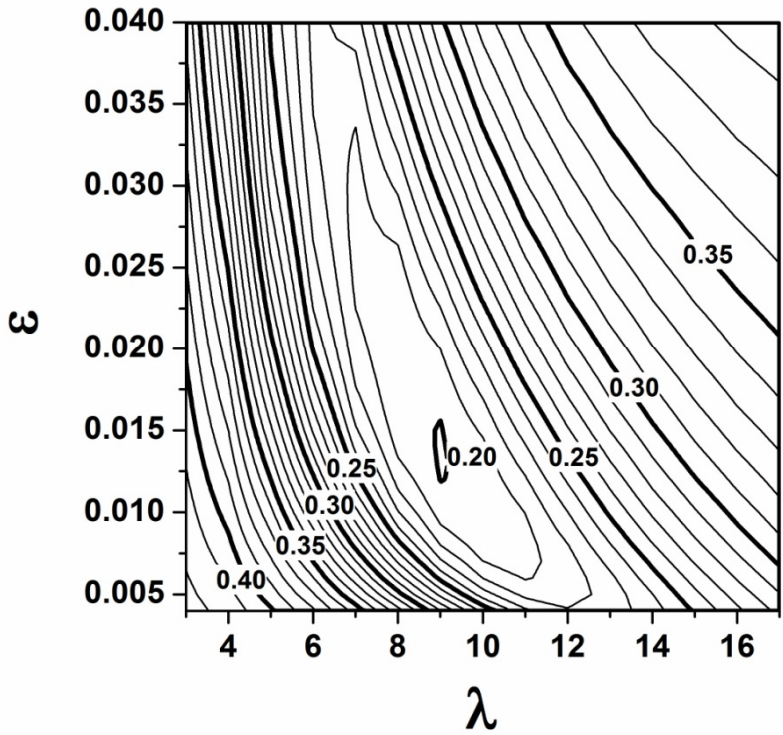

Figure 16: Contour plot for the degree of agreement $R$ [11] between measured and simulated ionisation cluster size distributions as a function of expectation value $\lambda$ and probability $\varepsilon$ of the distribution of the background of additional ionisations for the average of all data sets obtained with helium ions and carbon ions in 1.2 mbar $C_3H_8$.

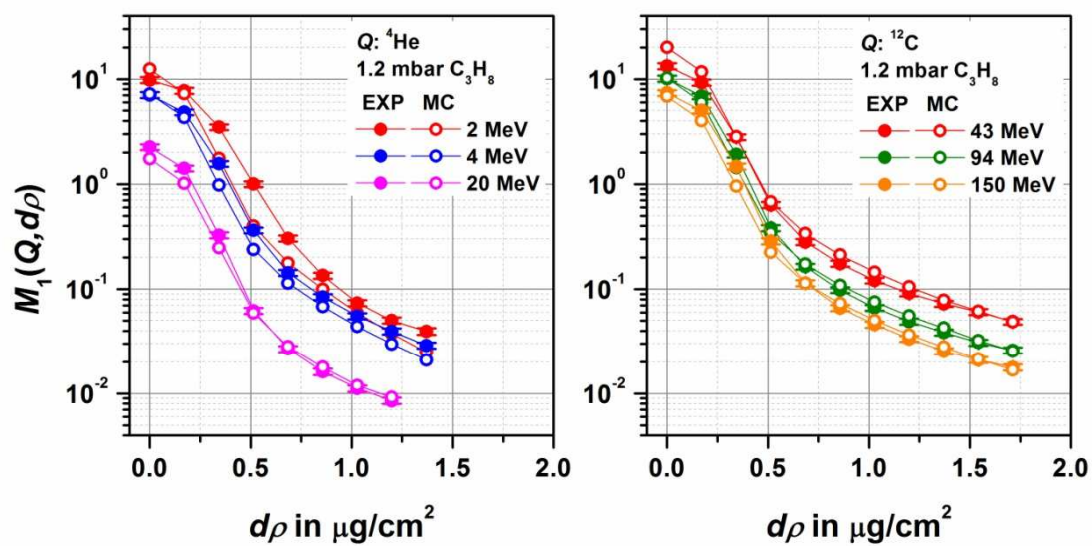

Figure 17: Comparison of the mean ionisation cluster sizes $M_1(Q,d\rho)$ obtained from measured and simulated background corrected ionisation cluster size distributions for helium ions (left) and carbon ions (right).

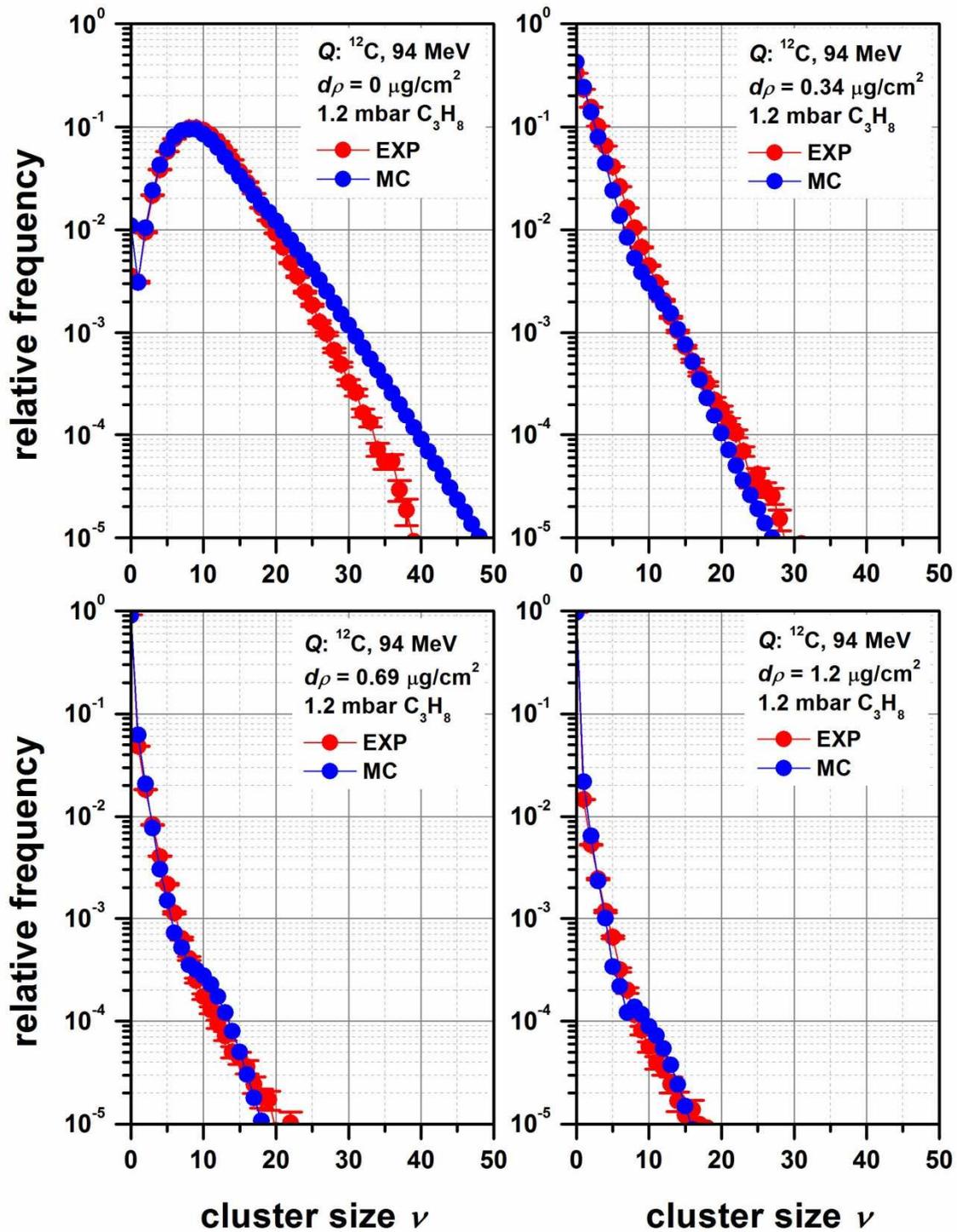

Figure 18: Comparison of measured and simulated background corrected ionisation cluster size distributions for selected values of the impact parameters $d\rho$ for 94 MeV carbon ions in 1.2 mbar $C_3H_8$.

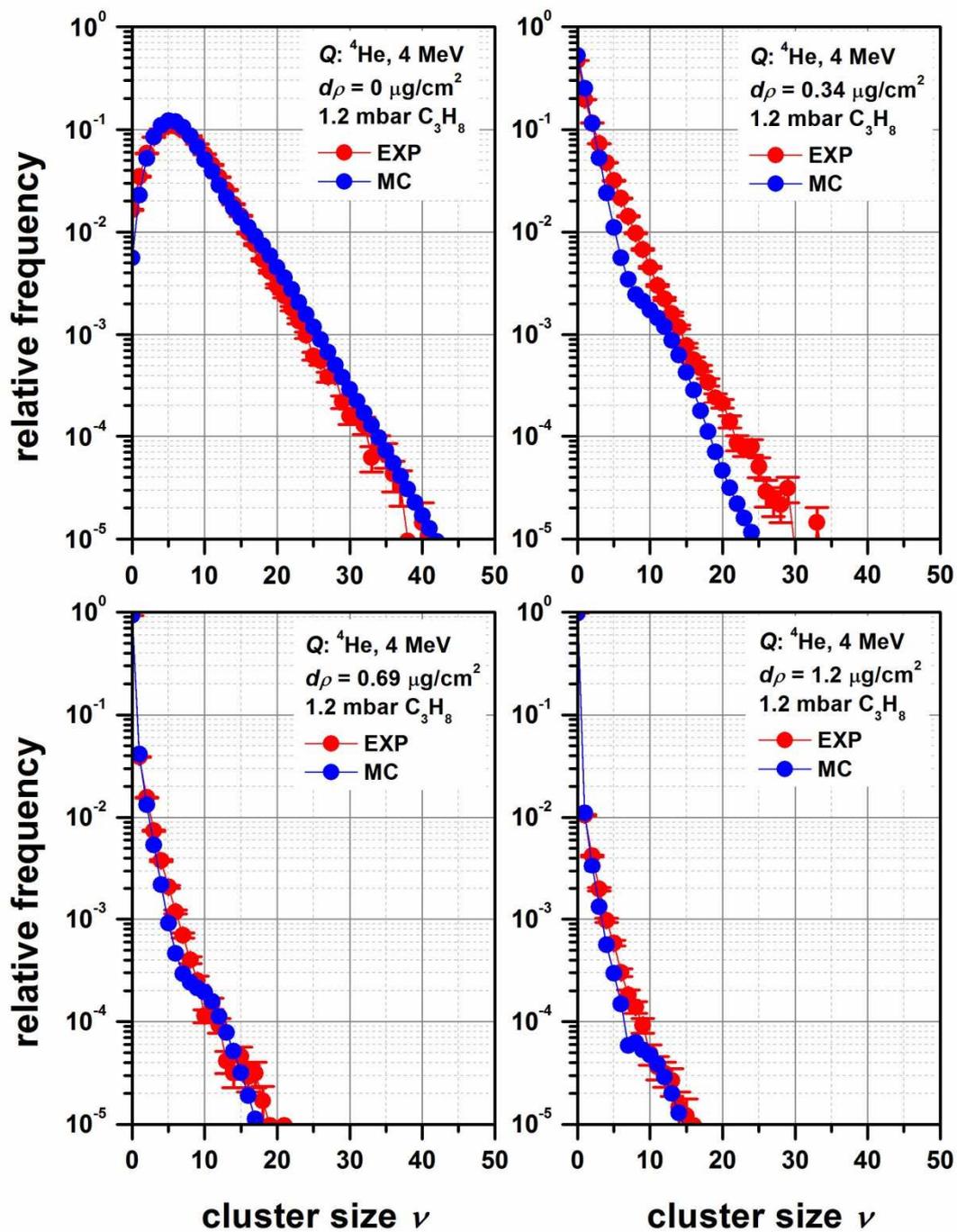

Figure 19: Comparison of measured and simulated background corrected ionisation cluster size distributions for selected values of the impact parameters $d\rho$ for 4 MeV helium ions in 1.2 mbar $C_3H_8$.

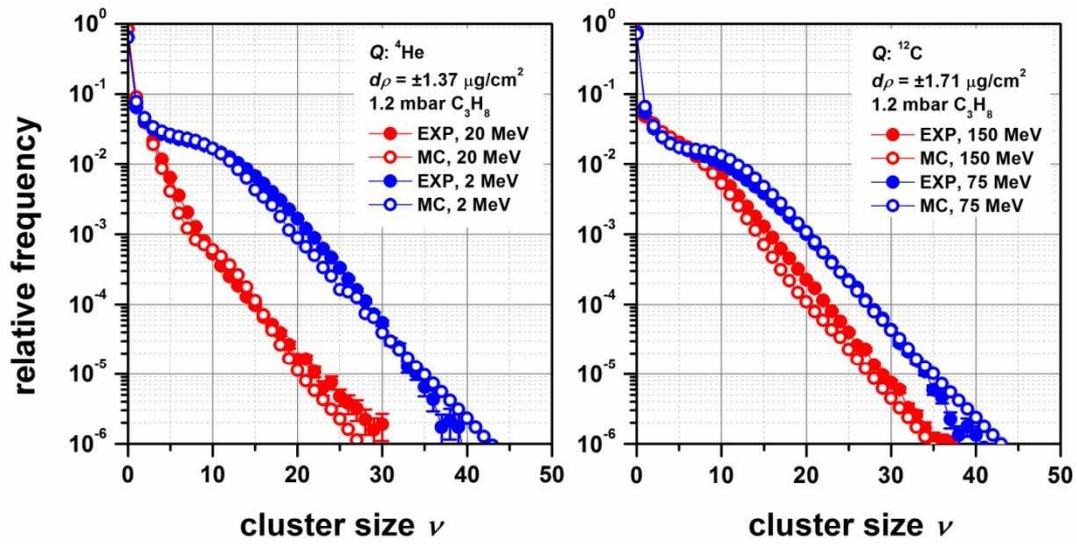

Figure 20: Comparison of measured and simulated background corrected ionisation cluster size distributions for broad beam geometry with the impact parameter ranging between $d\rho = \pm 1.37$ μg/cm² measured for helium ions (left) and between $d\rho = \pm 1.71$ μg/cm² for carbon ions (right) in 1.2 mbar $C_3H_8$.